\documentclass{aa}  
\usepackage{float}
\usepackage{graphicx}
\usepackage[dvipsnames]{xcolor}
\usepackage{txfonts}

\usepackage[pdfencoding=auto,psdextra]{hyperref}
\hypersetup{
    colorlinks=true,
    linkcolor=blue,
    filecolor=magenta,      
    urlcolor=blue,
    citecolor=blue
}
\usepackage{cleveref}
\crefname{chapter}{Chap.}{Chaps.}
\crefname{section}{Sect.}{Sects.}
\crefname{figure}{Fig.}{Figs.}
\Crefname{chapter}{Chapter}{Chapters}
\Crefname{section}{Section}{Sections}
\Crefname{figure}{Figure}{Figures}

\title{Emulating the Non-Linear Matter Power-Spectrum in Mixed Axion Dark Matter Models}

\begin{document} 
   \author{Dennis Fremstad
        \and 
        Hans A. Winther
    }
   \institute{Institute of Theoretical Astrophysics, University of Oslo,
PO Box 1029, Blindern 0315, Oslo, Norway 
    }
 
  \abstract
  {
  In order to constrain ultra light dark matter models with current and near future weak lensing surveys we need the predictions for the non-linear dark matter power-spectrum. This is commonly extracted from numerical simulations or from using semi-analytical methods. For ultra light dark matter models such numerical simulations are often very expensive due to the need of having a very low force-resolution often limiting them to very small simulation boxes which do not contain very large scales. 
  In this work we take a different approach by relying on fast, approximate $N$-body simulations. In these simulations, axion physics are only included in the initial conditions, allowing us to run a large number of simulations with varying axion and cosmological parameters.
  From our simulation suite we use machine learning tools to create an emulator for the ratio of the dark matter power-spectrum in mixed axion models -- models where dark matter is a combination of CDM and axion -- to that of $\Lambda$CDM. The resulting emulator only needs to be combined with existing emulators for $\Lambda$CDM to be able to be used in parameter constraints. We compare the emulator to semi-analytical methods, but a more thorough test to full simulations to verify the true accuracy of this approach is not possible at the present time and is left for future work. 
  }


   \titlerunning{Emulating the Non-Linear Matter Power-Spectrum in Mixed Axion Dark Matter Models}
   \authorrunning{Dennis Fremstad and Hans A. Winther}
   \maketitle


\section{Introduction}
Cold Dark Matter (CDM), which is thought to be a weakly interacting massive particle (WIMP), is the current leading candidate for dark matter. 
It successfully describes a number of observations, such as the internal dynamics of galaxy clusters \citep{Zwicky:1937, Clowe:2006}, the rotation curve of spiral galaxies \citep{Rubin:1980, Persic:1996}, and the weak gravitational lensing effects produced by large-scale matter structures \citep{Mateo:1998, Heymans:2013, Planck:2015}. 
However, the search for cold dark matter through indirect astronomical measurements \citep{Albert:2017}, direct laboratory experiments \citep{Danninger:2017} and through high-energy collider experiments \citep{Buonaura:2017} have not yielded any conclusive evidence for dark matter in the GeV range. 


Additionally, some discrepancies between observations and simulations of CDM around $k\sim 10$ kpc scales may suggest that the CDM model is insufficient when it comes to describing sub-halo structures. One such discrepancy, which has become known as the \textit{cusp-core} problem \citep{Oh:2011,deBlok:2009}, is the observation of relatively flat density profiles towards the center of dark matter halos \citep{Walker:2011}, while $N$-body simulations suggest an increasing profile \citep{Navarro:1996}. Another discrepancy is the missing satellite problem \citep{Klypin:1999, Moore:1999}, where $N$-body simulations predict a much higher abundance of satellite galaxies than what is found in observations. While these issues could be explained by an imperfect implementation of baryonic physics in numerical simulations \citep{Macci:2012}, some studies claim that these issues persist even when accounting for small scale baryonic physics \citep{Pawlowski:2015, Sawala:2014}. This sparked the search for alternative dark matter models that behave differently from CDM on galactic scales, while still being successful on cosmological scales. 

One such model is the axion, an extremely light boson originally proposed by string theory \citep{Arvanitaki:2010}. 
This form of dark matter belongs to a suite of wave-like dark matter also known as fuzzy dark matter (or $\psi$DM). Axions form a non-relativistic Bose-Einstein condensate, which means that the uncertainty principle leads to a self-interacting pressure. This pressure counteracts gravity on scales smaller than the Jeans scale \citep{Schive:2014dra}. As a result, on these smaller scales, perturbations oscillate rather than grow. However, on larger scales, axions behave in the same way as CDM \citep{Schive:2016}. This makes them an attractive dark matter model, as they might alleviate discrepancies on galactic scales, without altering large scale structure formation. 


The evolution of the axions follow the Schr\"{o}dinger-Poisson equations, which can be solved using high-resolution Adaptive Mesh Refinement (AMR) algorithms \citep{Schive:2018}. 
While this approach leads to impressive results in terms of resolution \citep{Schive:2014dra, Woo:2009}, it is highly resource-intensive, making it impractical for including a full hydrodynamic description of gas and star formation in cosmologically representative simulation domains.
One way to reduce the computational demand is by incorporating the dynamical effect of quantum pressure via smoothed particle hydrodynamics (SPH), allowing for less intensive simulations without sacrificing cosmological results \citep{Nori:2018hud}.
Another approach is to use standard $N$-body codes and include the effects of axions only in the initial conditions, thereby neglecting axion physics in the simulations \citep{VidI:2017, Schive:2016, Armengaud:2017}.

The standard approach to inferring cosmological parameters, including those related to axions, is through Bayesian techniques. 
However, approaches that rely on generating numerous accurate theoretical predictions, such as a Markov chain Monte Carlo (MCMC) approach, quickly become computationally expensive due to the need for many $\it N$-body simulations. 
A promising approach that may alleviate the numerical load is the use of emulators, which can model the relation between cosmological parameters and observables such as the power spectrum. Some examples of cosmological emulators include \textsc{EuclidEmulator1}, \textsc{EuclidEmulator2} \citep{Euclidemu:2019, Euclidemu:2021}, \textsc{FrankenEmu} \citep{Heitmann:2013}, \textsc{Dark quest} \citep{Nishimichi:2019}, and the \textsc{Bacco} emulator \citep{Angulo:2020vky}.


In this paper, we use a fast approximate $N$-body simulation, known as the \textsc{COLA} method, to simulate the axions as standard CDM particles, where the axion physics come into play through the initial conditions alone. These initial conditions are generated using \textsc{axionCAMB}, where the axion mass and the axion abundance, that is the fraction of the total dark matter energy budget that consists of axions, are free parameters.
Using this approach, we generate a data-set which we use to train an emulator that quickly estimates the non-linear matter power-spectrum with axions.

\section{Theory}
In this section we review the basics of mixed axion models from its formulation, background evolution, perturbations and simulations. We also discuss semi-analytical methods for predicting the power-spectrum.

\subsection{Background cosmology}

The action for an axion scalar field is
\begin{equation}\label{eq:klein-gordon}
    S = \int d^4x\sqrt{-g}\left[\frac{1}{2}g^{\mu\nu}\partial_\mu \phi\partial_\nu \phi - \frac{1}{2}m_{\rm ax}\phi^2\right],
\end{equation}
and gives rise to equation of motion (the Klein-Gordon equation) 
\begin{equation}\label{eq:klein-gordon}
    \square \phi - m_{\rm ax}^2 \phi = 0,
\end{equation}
where
\begin{equation}
    \square = \frac{1}{\sqrt{-g}} \partial_\mu \left(\sqrt{-g}g^{\mu\nu}\partial_\nu\right),
\end{equation} 
is the d'Alembert operator. At zeroth order, equation \ref{eq:klein-gordon} describes the cosmological background of the axion field, and takes the form of a damped harmonic oscillator \citep{Marsh:2015xka}. The equation is given by
\begin{equation}
    \ddot{\phi}_0 + 3H\dot{\phi}_0 + m_{\rm ax}^2 \phi_0 = 0.
\end{equation}
When $3H \gg m_{\rm ax}$, the field is over-damped and essentially "frozen". This causes the axions to fulfill the slow-roll condition, making them behave similar to dark energy. Once $m_{\rm ax} \sim 3H$ the field begins to oscillate, and the axions begin to behave as dark matter.

To see this more clearly, in a matter or radiation dominated Universe $a\propto a^n$ and the solution is given by
\begin{equation}
    \phi = a^{-3/2} (t/t_{\rm ini})^{1/2}[A J_\ell(mt) + B Y_\ell(mt)],
\end{equation}
where $J_m,Y_m$ are Bessel-functions, $t_{\rm ini}$ is the initial time, $\ell = (3n-1)/2$ and $A,B$ set by the initial conditions.  

The energy density and pressure of the axion field are given by
\begin{align}
    \rho &= \frac{1}{2}\dot{\phi}^2 + \frac{1}{2}m^2\phi^2,\\
    P &= \frac{1}{2}\dot{\phi}^2 - \frac{1}{2}m^2\phi^2.
\end{align}
Averaged over time the solution above, for $m_{\rm ax} \sim 3H$, leads to $\rho \propto a^{-3}$ and $P \approx 0$ and the axion behaves as dark matter. In the late Universe and for the axion masses of interest for us ($10^{-22}$ eV - $10^{-28}$ eV) the background evolution will be practically the same as that for $\Lambda$CDM.


\subsection{Linear perturbation theory}

To get the linear perturbation theory approximation of the evolution of the axion scalar field, one can apply a perturbation $\delta \phi$ to the axion field $\phi = \phi_0 + \delta \phi$. With the axion overdensity, given by $\delta_{\rm ax} = 1 + \frac{\delta \rho}{\bar{\rho}}$ and using the Klein Gordon equation Eq.~(\ref{eq:klein-gordon}) one can arrive at the equations of motion for the axion overdensity \citep{Vogt_2023}
\begin{equation}
    \delta_{\rm ax}' = kv_{\rm ax} - 3 \Phi',
\end{equation}
\begin{equation}
    v_{\rm ax}' = -\mathcal{H} v_{\rm ax} - c_s^2k\delta_{\rm ax} - k\Psi.
\end{equation}
Here, the derivatives are with respect to conformal time $\tau$ (d$t = a(t)$d$\tau$), $c_s$ is the sound of speed of the axions, given by 
\begin{equation}
    c_s^2 = \frac{\frac{k^2}{4 m^2_{\rm ax}a^2}}{1 + \frac{k^2}{4m^2_{\rm ax}a^2}},
\end{equation}
and $\Phi,\Psi$ are the perturbations to the metric. For $k\gg m^2$ the sound-speed goes to unity which prevents clustering on the smallest scales while on the largest scales $c_s^2 \to 0$ which gives the same behavior as cold dark matter.

\subsection{Numerical simulations of axions}

In the non-relativistic limit, we can expand \cref{eq:klein-gordon} in terms of a complex field $\psi$ using the Wentzel-Kramers-Brillouin (WKB) ansatz, 
\begin{equation}
    \phi = \frac{1}{\sqrt{2m_\text{ax}^2}}\left(\psi e^{im_\text{ax}t} + \psi^* e^{-im_\text{ax}t}\right).
\end{equation}
This factors out some of the oscillations ($e^{\pm i m_{\rm ax}t}$) experienced by the axions. The complex field $\psi$ then follows the Schr\"{o}dinger equation
\begin{equation}\label{eq:schrodinger}
    i \frac{\partial}{\partial t}\psi + \frac{3Hi}{2}\psi = -\frac{1}{2m_\text{ax}a^2} \nabla^2\psi + m_\text{ax}\Phi_N \psi,
\end{equation}
with $\Phi_N$ as the Newtonian potential, given by the Poisson equation 
\begin{equation} \label{eq:newtonian-potential}
    \nabla^2\Phi_N = 4 \pi G a^2 \left[m_\text{ax}(|\psi|^2 - \langle|\psi|\rangle^2) + \delta\rho_f \right].
\end{equation}
In Eq.~(\ref{eq:newtonian-potential}), $|\psi|^2 - \langle |\psi|\rangle^2$ serves as the overdensity $\delta\rho_{\rm ax}$ for the Axion scalar field, while $\delta \rho_{f}$ describes the overdensity of CDM and baryons.

An alternative description of the axion fluid is provided by the so-called Madelung formulation of the Schr\"{o}dinger equation. Defining $\psi = \sqrt{\rho/m}e^{i\theta}$ and, ${\bf v} = \nabla\theta/m$ then Eq.~(\ref{eq:schrodinger}) can be written as the fluid equations
\begin{align}
    &\frac{\partial \rho}{\partial t} + \nabla\cdot(\rho {\bf v}) = 0,\\
    &\frac{\partial {\bf v}}{\partial t} + ({\bf v}\cdot \nabla){\bf v} = -\nabla\Phi_N - \nabla Q,
\end{align}
where $Q = -1/(2m^2)\cdot (\nabla^2\sqrt{\rho})/\sqrt{\rho}$. This is the well-known continuity and Euler equation (for an irrotational fluid $\nabla\times {\bf v} = 0$) where $Q$ acts as a pressure term, commonly called the quantum pressure. Even though the equations describe an irrotational fluid, in the region where $\rho\to 0$ the phase can develop discontinuities which again can generate vorticity in the field. 

These different formulations mentioned above gives rise to different approaches of simulating the axion:
\begin{itemize}
    \item The operator splitting technique with finite differencing:
    \begin{equation}
    \psi(x,t+\Delta t) = e^{-im\Phi_N \Delta t} e^{\frac{i}{2m}\Delta t \nabla^2}\psi(x,t).
    \end{equation}
    The last (kinetic) operator is computed by Taylor expanding it and evaluating the resulting terms $\nabla^{2n}\psi$ using finite difference. The main advantage of this method is that it generalizes to a non-uniform mesh. This was the original approach of \cite{Schive:2014dra} and was later implemented in the \texttt{SCALAR} code \citep{Mina:2019ekb}.
    \item Pseudo-spectral methods: This is similar to the method above, but where the kinetic operator is evaluated in Fourier space, i.e.
    \begin{equation}
    e^{\frac{i}{2m}\Delta t \nabla^2}\psi(x,t) = \mathcal{F}^{-1}[e^{-\frac{i}{2m}\Delta t k^2}\mathcal{F}[\psi(x,t)]].
    \end{equation}
    The advantage of this method is that it is very accurate, but it requires us to have a full grid. To get high resolution, the box size either needs to be small or the number of grid-cells needs to be very large. This is the approach taken by, e.g., \cite{May:2021wwp,May:2022gus}.
    \item Smoothed particle hydrodynamics: This uses the Madelung formulation and the axion is modeled using tracer particles, from which one can estimate the density and quantum pressure to evolve the equations. This approach is taken by \texttt{AX-GADGET} \citep{Nori:2018hud}.
    \item Hybrid methods: One can use the pseudo-spectral algorithm on the root grid and finite differencing on adaptively refined grids. This is the approach taken by \texttt{AxioNyx} \citep{Schwabe:2020eac}.
\end{itemize}
The main issue with most of these methods is that since the Schr\"{o}dinger equation is first order in time and second order in space, the stability is only guaranteed when $\Delta t \lesssim (\Delta x)^2$. In addition to this the spatial resolution $\Delta x$ should satisfy $\Delta x \lesssim \lambda_{\rm dB}$, the de-Brouigle wavelength of the axion, which typically is very small $\mathcal{O}(\text{kpc})$. This means that we need very fine grids and very small time steps, making the simulations very costly. This issue is avoided for the SPH approach, as there is no underlying mesh, but it still requires fine time-steps. However, for this method, there is also the question if it really can produce all the wave-like behavior we get when solving the Schr\"{o}dinger equation. 

For this reason, most of the simulations done so far have restricted themselves to very small simulation boxes $B = \mathcal{O}(1-15~\text{Mpc})$. Many of these simulations have also not been simulated all the way to redshift $z=0$, but rather ending at a higher redshift.

The approach we take in this paper is the most naive one: we simply include the effect of the axion, as predicted by linear perturbation theory, only in the initial conditions. This ignores the effect of the quantum pressure, but do allow us to simulate large cosmological boxes.

It is hard to properly assess the accuracy of our approach, as there are few big-box simulations available in the literature. For simulations using the Madelung formulation there have been simulations (with a box size of $10-15$ Mpc$/h$) performed with and without including the quantum pressure and a very good agreement was found when in comes to the matter power-spectrum (see e.g. Fig. 9 in \citep{Nori:2018hud}). In \cite{Veltmaat:2016rxo} it was shown that cosmological simulations (with a box size of $2$ Mpc$/h$)  including the quantum pressure gives rise to a maximum relative difference of $10\%$ in the power spectrum near the quantum Jeans length. All these simulations were for an axion abundance, $f_{\rm ax} \equiv \Omega_{\rm ax}/(\Omega_{\rm ax} + \Omega_{\rm CDM})$, of $f_{\rm ax} = 1$ and we would expect differences to diminish with decreasing axion abundance. 

\subsection{The halo model and the \textsc{axionHMCode} prediction}

The halo model \citep{Seljak:2000gq} assumes that all the matter in the Universe is located inside halos. With this assumption the 2-point correlation function can then be written analytically as the sum of two terms: first we have the 2-halo term, $P_{2h}$, which takes into account the correlations between mass-elements in two different halos, and then we have the 1-halo term, $P_{1h}$, which takes into account correlations between mass-elements within the same halo. For the power-spectrum this gives us
\begin{align}
    P(k) &= P_{\rm 1h}(k) + P_{\rm 2h}(k),\\
    P_{\rm 2h}(k) &= P_{\rm lin}(k)\left[\frac{1}{\overline{\rho}_m}\int \frac{dn}{d\log M} b(M)y(k,M) dM\right]^2,\\
    P_{\rm 1h}(k) &= \frac{1}{(2\pi)^3}\int \frac{dM}{\overline{\rho}_m^2}\frac{dn}{d\log M} M y(k,M)^2, 
\end{align}
where $P_{\rm lin}(k)$ is the linear dark matter power-spectrum, $b(M)$ is the halo-bias, $dn/d\log M$ is the halo mass-function and $y(k,M)$ is the Fourier transform of the halo density profile $\rho(r, M)$. The 2-halo term dominates the prediction on large scales (small $k$) and the 1-halo term dominates on small-scales.

The ingredients needed to evaluate the halo model are the density profile $\rho(M)$, the mass-function $dn/d\log M$ and the bias $b(M)$. The latter two are for $\Lambda$CDM often taken to be the analytical fit made by Sheth-Tormen \citep{Sheth:2001} and for the former it is most often taken to be a NFW profile with a mass-concentration relation $c(M)$ derived from simulations.

For axions all these ingredients are expected to change. First of all, the density profile will in general be cored and the mass-function will have a cut-off for small masses. Another complication that arises when we consider mixed dark matter models, where only part of the DM budget is in the form of axions and the rest in the form of CDM. This leads to a prediction
\begin{align}
    P(k) &= (1-f_{\rm ax})^2P_{\rm CDM}(k)^2 + P_{\rm axions}^2 f_{\rm ax}^2 + 2f_{\rm a}(1-f_{\rm ax})P_{\rm CDM \times axion}(k),
\end{align}
where each of the terms above can be computed in the halo-model, and the full equations can be found in \cite{Vogt_2023}. This is implemented in the \textsc{axionHMCode} package\footnote{\url{https://github.com/SophieMLV/axionHMcode}} \citep{Vogt_2023} and includes improvements to the model by calibrating it to axion simulations as performed in \cite{Dome:2024}.

\textsc{axionHMCode} takes into account the cut-off in the halo mass-function and uses a solitonic core for the axion density profile (derived from the simulations of \cite{Schive:2014dra}):
\begin{equation}
    \rho_c (r) = \frac{1.9(1+z)}{(1 + 9.1\times 10^{-2} (r/r_c)^2)^8}\left(\frac{r_c}{{\rm kpc}}\right)^{-4} \left( \frac{m_{\rm ax}}{10^{-23} {\rm eV}}\right)^{-2} M_{\odot} {\rm pc}^{-3},
\end{equation}
where $r_c$ is the core radius defined as where the density drops to one half of the central density, and has been found to be well approximated by \citep{Schive:2014dra}
\begin{equation}
    r_c = 1.6(1+z)^{-1/2} \left(\frac{m_{\rm ax}}{10^{-22}{\rm eV}} \right)^{-1} \left(\frac{\Delta_v(z)}{\Delta_v(0)}\right)^{-1/6}\left(\frac{M_{\rm h}}{10^9 M_{\odot}}\right)^{-1/3} {\rm kpc},
\end{equation}
where $\Delta_v(z)$ is the halo virial overdensity. The \textsc{axionHMCode} approach includes all the physics we expect to be present in mixed axion DM models, except for the presence of interference patterns. However, these are small-scale effects and are only expected to lead to modifications on very small scales beyond what we are interested in modeling with the power-spectrum. In the absence of simulations with large boxes to compare to, we will use this package as the main comparison throughout this paper.

\section{Simulation setup} 

We use \textsc{axionCAMB}\footnote{\url{https://github.com/dgrin1/axionCAMB}}  \citep{Hlozek_2015,2022ascl.soft03026G}, a modified version of the Einstein-Boltzmann solver \textsc{CAMB}\footnote{\url{https://github.com/cmbant/CAMB}} which includes Axion physics \citep{Lewis_2000}, to generate the linear matter power-spectrum $P(k,z=0)$ for any combination of cosmological parameters, axion mass and axion abundance. Combined with the common back-scaling technique \citep[see e.g.][]{Fidler:2017ebh}, we are able to use the generated power-spectrum from \textsc{axionCAMB} to formulate initial conditions in a universe with mixed dark matter. The redshift of the initial conditions used in this work is set to $z_{\rm ini} = 30$, where essentially all scales of interest are linear and well described by the linear matter power-spectrum. These initial conditions can then be evolved by a fast approximate $N$-body simulation to generate a snapshot of the universe at any given redshift. We let the axions be described by usual cold dark matter particles and evolve it forward in time using a particle-mesh $N$-body code. Thus, with our approach, the effect of axions is solely included in the initial conditions.

\begin{figure}
    \centering
    \includegraphics[width=\linewidth]{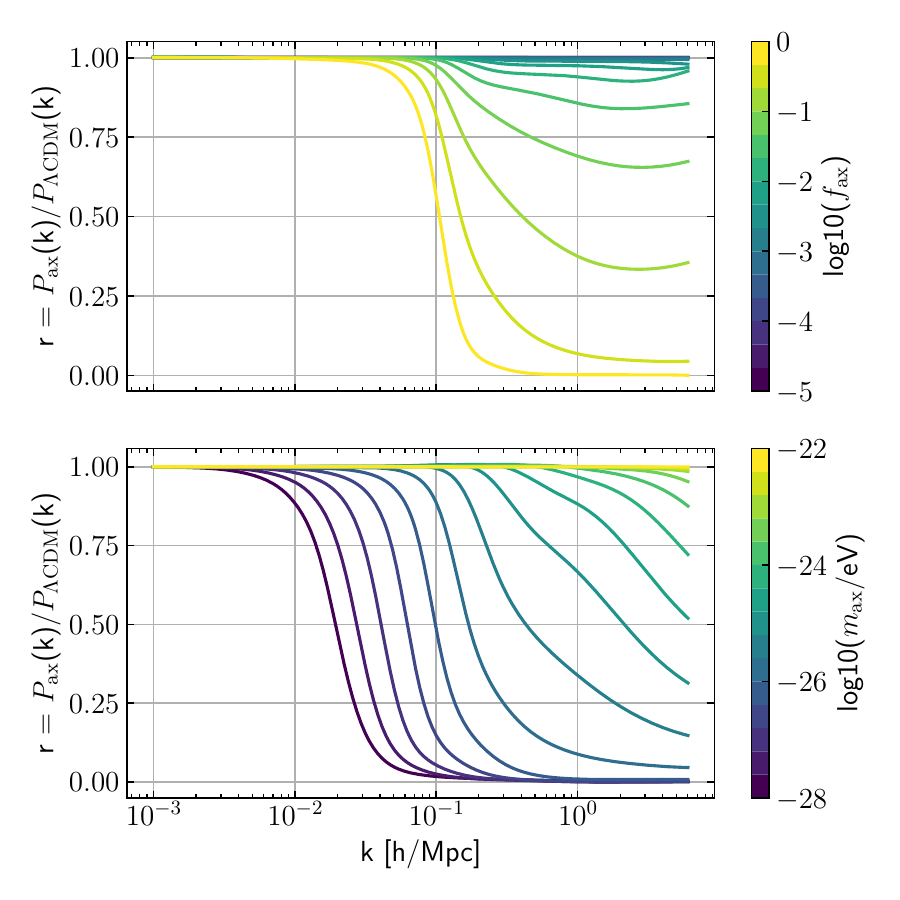}
    \caption{
    An example of the non-linear power-spectrum ratio $r$ for varying axion abundance (top panel) and axion mass (bottom panel). We use axion mass $m_{\rm ax} = 10^{-26}$ eV in the top panel, and axion abundance $f_{\rm ax} = 0.5$ in the bottom panel. The data in this figure was generated using the trained axion emulator described in this paper.}
    \label{fig:axion_dependency}
\end{figure}

To speed up the simulations, we use the \textsc{COLA} method \citep{Tassev}, which involves transforming the $N$-body equations of motion into the frame following the evolution predicted by Lagrangian Perturbation Theory (LPT). In this frame, the equations of motion takes on the following form,
\begin{align}
    \partial_t^2 \boldsymbol x_{\rm res} = - \nabla \Phi - \partial_t^2 \boldsymbol x_{\rm LPT},\\
    \nabla^2\Phi = 4\pi Ga^2 \delta \rho,
\end{align}
where we solve for $\boldsymbol x_{\rm res} = \boldsymbol x - \boldsymbol x_{\rm LPT}$: the correction added to the LPT solution $\boldsymbol x_{\rm LPT}$. Thus, the position of the particles are given by the sum of the position of the particles as found with LPT and the simulated corrections to the particle positions. 
The advantage of this is that it allows us to take larger time-steps than usual while still maintaining accuracy on the largest scales. See e.g. \cite{Winther:2017jof} for more details about the COLA method. The drawback of using this simulation method is that COLA, and particle-mesh simulations in general, suffer from lack of resolution on small scales. However, some of this can be factored out by looking at the ratio between the matter power-spectrum with mixed axion dark matter and $\Lambda$CDM,
\begin{equation} \label{eq:ratio}
    r = \frac{P_{\rm ax}}{P_{\Lambda \rm CDM}},
\end{equation}
both simulated with the COLA method. This will also factor out much of the dependency on the cosmological parameters, and has a smooth shape, which makes it easier to emulate. 
We present an example of this ratio in \cref{fig:axion_dependency}. 


\subsection{Testing the \textsc{COLA} setup}

Hundreds of simulations are required to generate a training data-set that probes the effect of the axion mass and abundance, and the cosmological parameters. For this reason, we look to establish an optimal simulation setup that is both cost-efficient and accurate. To do this, we run a pair of $\Lambda$CDM and mixed dark matter simulations for a fiducial set of cosmological and simulation parameters to find the power-spectrum ratio. By generating a new pair of simulations with one of the simulation parameters altered, and looking at the relative difference between this and the fiducial simulation run, we can study the effect of each individual parameter on the power-spectrum ratio. 

\begin{figure}
    \centering
    \includegraphics[width=\linewidth]{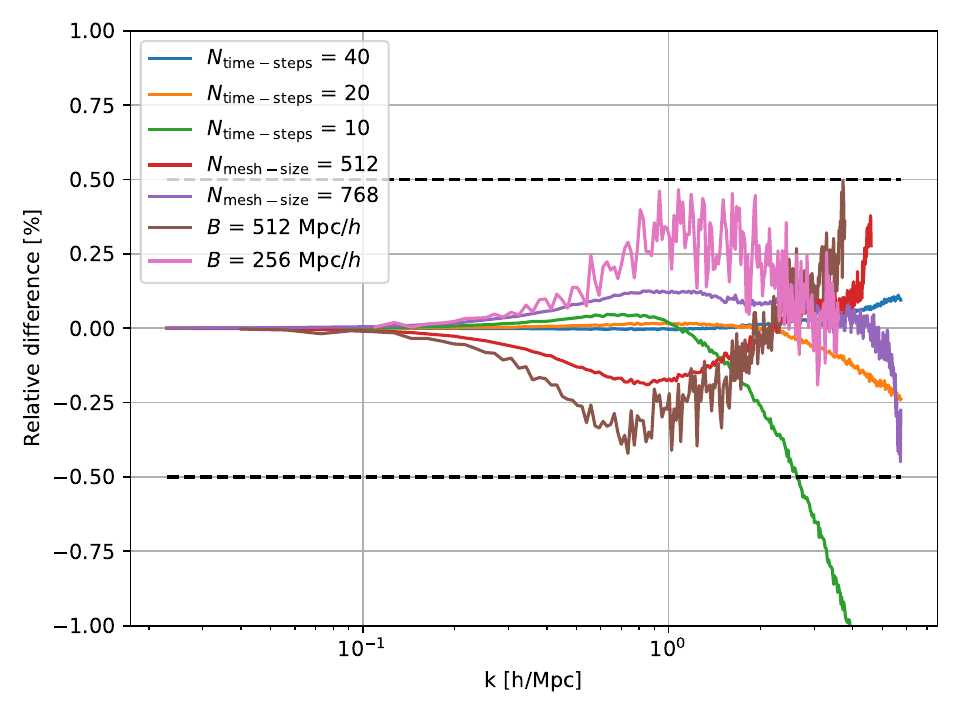}
    \caption{Tests of how the power-spectrum ratio $r$ changes with varying simulation parameters: box-size, number of time-steps and grid-size (force resolution). A control simulation with parameters $N_{\rm time-steps} = 30$, $N_{\rm mesh-size} = 640$ and $B = 350$ Mpc/h was used.}
    \label{fig:cola_test}
\end{figure}

We test the number of time-steps, the number of grids (which is set to be equal to the number of particles) and the box size, and plot the result in \cref{fig:cola_test}. This test was carried out using our fiducial cosmology, but we have also performed tests to check that we have the same qualitative behavior for other cosmologies (i.e. for a few different axion masses and fractions).
Our fiducial simulation setup uses $30$ time-steps, $640$ mesh grid-points and has a box-size of $350$ Mpc/h. We find that as long as the number of time-steps is greater than $\sim 20$, we already have sub-percent convergence. Perturbing the other parameters yields a $\sim 0.5 \%$ difference in the power-spectrum ratio. For this reason, we run all our simulations using the fiducial setup. 
Initial conditions are created using 2LPT at a starting redshift of $z=30$.


\begin{figure}
    \centering
    \includegraphics[width=\linewidth]{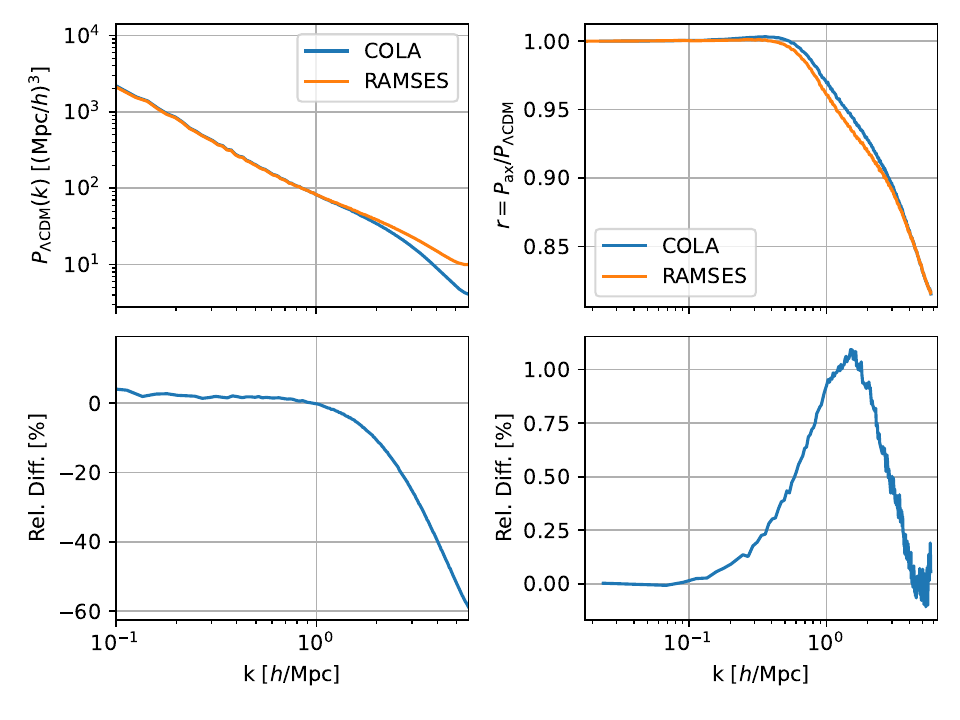}
    \caption{Comparison of the power-spectrum ratio $r$ as calculated with COLA to what is found with RAMSES. The top left panel shows the raw power-spectrum found using \textsc{COLA} and \textsc{RAMSES}, while the bottom left panel shows the relative difference between the two. The top right panel shows the ratio between the power-spectrum from $\Lambda$CDM and axions, and the bottom right panel shows the relative difference of these.}
    \label{fig:test_ramses}
\end{figure}

We also test the \textsc{COLA} method itself by comparing its results with what is found when evolving the initial power-spectrum with \textsc{RAMSES}. By comparing our approach with \textsc{RAMSES}, we are able to investigate the underlying error of the \textsc{COLA} method. The power-spectrum and power-spectrum ratio from both \textsc{COLA} and \textsc{RAMSES}, along with the relative difference between them, are plotted in \cref{fig:test_ramses}. Computing the power-spectrum with \textsc{COLA}, we find that we induce a relative difference of up to $\sim 60 \%$ when compared to \textsc{RAMSES}. At the same time, computing the power-spectrum ratio, we find that we induce a relative difference of only $\sim 1\%$ on small scales.



\subsection{Constructing the Latin Hypercube}

As previously mentioned, by emulating the power-spectrum ratio, we factor out some of the cosmological parameters. It is therefore important to distinguish between which parameters still has an effect on the power-spectrum ratio, in order to determine what parameters need to be included in the emulation. To accomplish this, we follow a similar procedure to the previous section, where we define a fiducial set of cosmological parameters, and perturb a single parameter to explore its effect. 

\begin{figure}
    \centering
    \includegraphics[width=\linewidth]{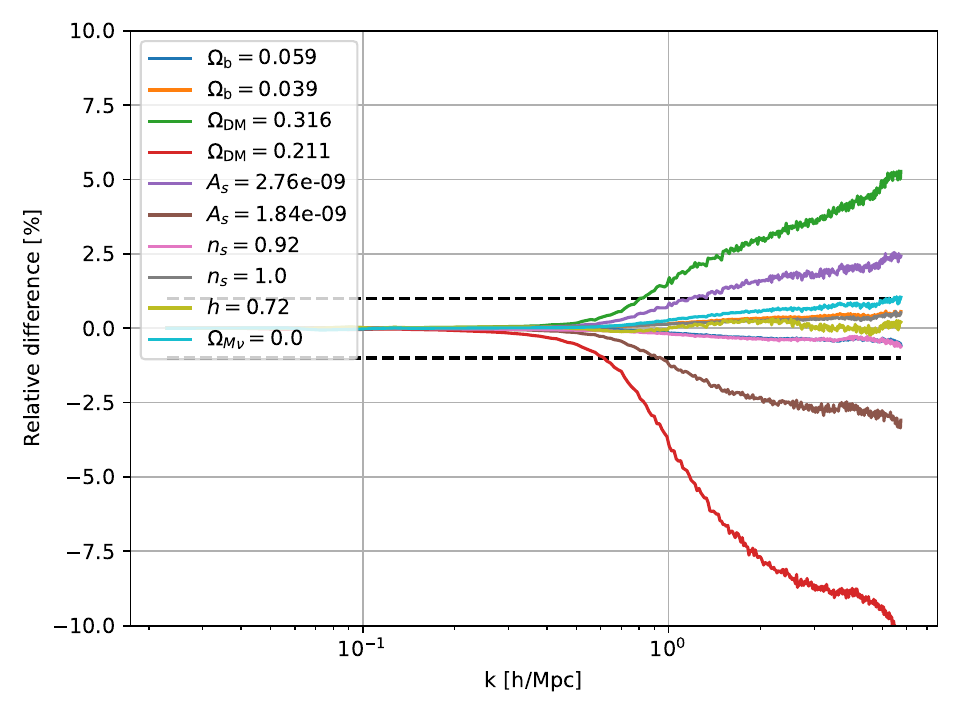}
    \caption{The variation of the power-spectrum ratio $r$ as function of cosmological parameters. The dashed line indicates a relative difference of $1\%$. Only $A_s$ and $\Omega_m$ have significant ($\gtrsim 1\%$) deviations. An axion abundance of $f_{\rm ax} =0.2$ and axion mass of $m_{\rm ax} = 10^{-24}$ eV was used.}
    \label{fig:test_cosmo}
\end{figure}

\cref{fig:test_cosmo} displays the effect of $\Omega_b$, $\Omega_{\rm DM}$, $A_s$, $n_s$, $h$ and $\Omega_{M\nu}$ on the power-spectrum ratio, where we have used the fiducial parameters
\begin{align*}
    \Omega_b &= 0.049, \\
    \Omega_{\rm DM} &= 0.2637, \\
    A_s &= 2.3 \cdot 10^{-9}, \\
    n_s &= 0.966, \\
    \Omega_{M\nu} &= 0.0048, \\
    h &= 0.6711. 
\end{align*}
We find that most of the parameters are almost completely factor out, and only contribute $\lesssim 1\%$ on small scales, however $A_s$ and $\Omega_m$ still exhibit a significant contribution to the power-spectrum ratio. Thus, we include this in the emulation. 
Furthermore, by ignoring massive neutrinos, which speeds up simulations by a factor $\sim 2$, we only induce an error of only $\sim 1\%$ on small scales, which is too small to be resolved by COLA due to its intrinsic small scale error. 
Though, the effect of all the ignored parameters are still induced implicitly since $P_{\rm ax} = r P_{\Lambda \rm CDM}$ and the latter factor carries parameter dependencies.


The range of the included cosmological parameters are determined by expanding the parameter range used in the \textsc{EuclidEmulator2} \citep{euclid_collaboration}. 
We choose to exclude a few of the highest redshift results from our \textsc{COLA} setup, leaving us with $28$ data-points linearly spaced in $1/(1+z)$ between $z=6.75$ and $z=0$.
The axion abundance is probed between $1$ and $0.001$, as we only want to look at mixed dark matter models. For the axion mass, we go from $10^{-28}$ eV, where the axion particles behave more similarly to dark energy \citep{Hlozek_2015}, to $10^{-22}$ eV, where we stop seeing the effect of the axions given our simulation setup. 
Thus, the final parameter space we end up with is given by:
\begin{align*}
    \log_{10}{m_{\rm ax}/{\rm eV}} &\in [-28, -22], \\
    f_\text{ax} &\in [0.001, 1], \\
    \log_{10}A_s &\in [-9, -8.52], \\
    \Omega_\text{DM} &\in [0.2, 0.4], \\
    z &\in [0, 6.75],
\end{align*}
where $\Omega_\text{CDM} = \Omega_\text{DM}(1-f_{\rm ax})$ (i.e. for $\Lambda$CDM $\Omega_\text{DM} \equiv \Omega_\text{CDM}$). 

We employ a simple wrapping scheme when creating the Latin hypercube sampling to ensure that our emulator has a sufficient amount of samples at the edges of the $m_{\rm ax}$ and $f_{\rm ax}$ parameter space. In short, we sample values slightly beyond the limits of our desired sampling, and wrap the values beyond these limits back to the edges, as given above. We use the following expanded sample limits 
\begin{align*}
    \log_{10} m_{\rm ax}/{\rm eV} &\in [-28.5, -21.5],  \\
    f_{\rm ax} &\in [-0.1, 1.1].
\end{align*}

\begin{figure}
    \centering
    \includegraphics[width=\linewidth]{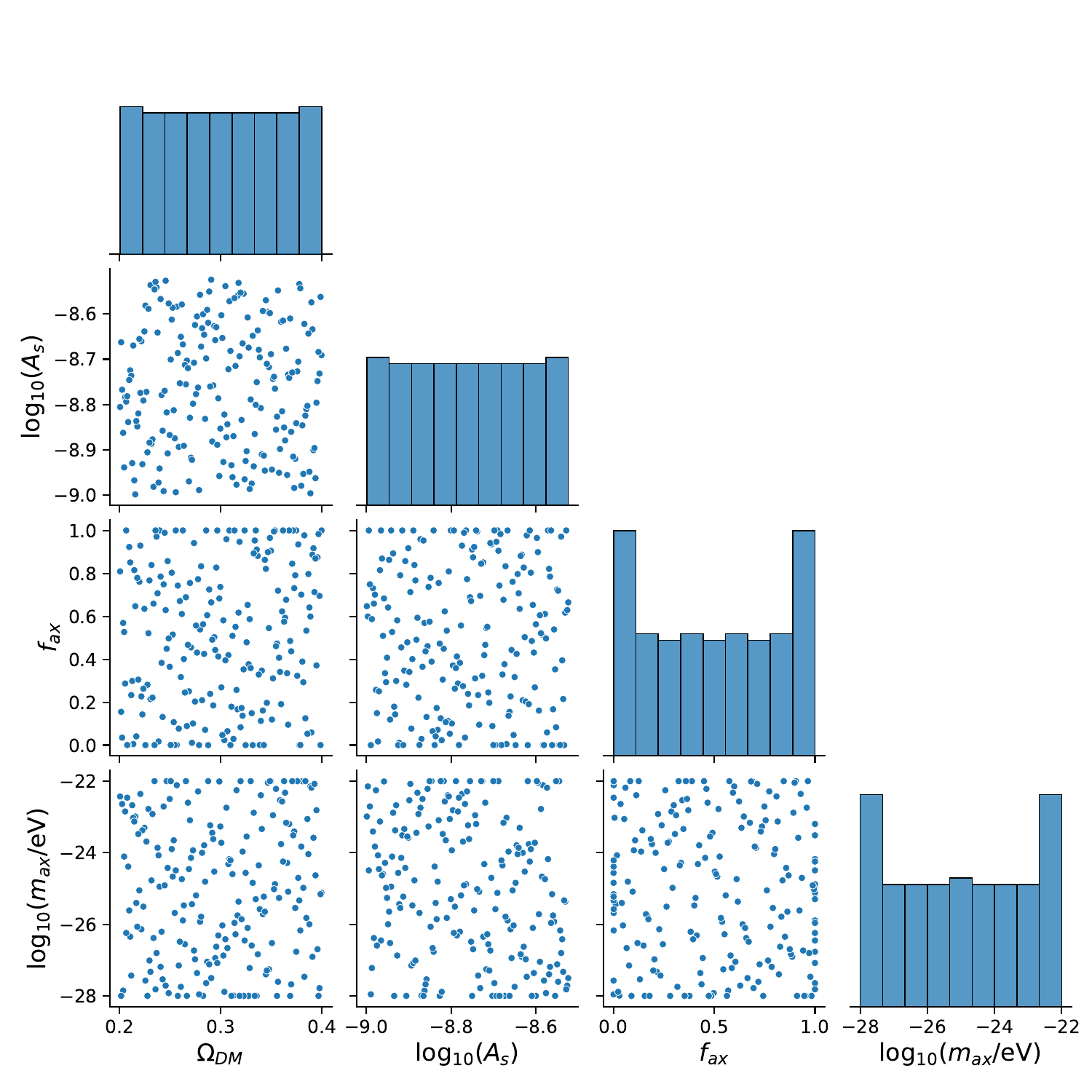}
    \caption{Distribution of samples in our Latin hypercube. We increase the sample density at some of the edges in order to improve the performance of the emulator in these regions. 
    }
    \label{fig:LHS}
\end{figure}

Using 200 data-points, we end up with the Latin hypercube sampling shown in \cref{fig:LHS}. Both $m_{\rm ax}$ and $A_s$ have been sampled logarithmically.


\subsection{The Final Setup}

In the final \textsc{COLA} simulation setup, we use a box size of $350$ Mpc/h, which gives us a force resolution of $\sim 0.5$ Mpc/h and a range from $k = 0.03$ h/Mpc to $k = 5.7$ h/Mpc. The simulations run from initial redshift $z_{\rm ini} = 30$ for $30$ time-steps. We use baryon density $\Omega_{\rm b} = 0.049$, Hubble parameter $h = 0.6711$ and spectral index $n_s = 0.966$. Neutrinos are ignored. 


To increase the range and accuracy of the emulator in the low-$k$ regime, we inject $256$ data-points from $k = 10^{-3}$ h/Mpc to $k = 0.02$ h/Mpc using \textsc{axionCAMB}, which is sensible since this regime is linear even at redshift $z=0$. Since the power-spectrum with axions always tends to a $\Lambda$CDM power-spectrum at low $k$, we get a plateau at $1$ for $P_{\rm ax}/P_{\Lambda \rm CDM}$ when we inject linear values. 
This helps stabilize the emulator during training.

\section{Emulator}

\subsection{Network Architecture \& Training}

We employ a simple Feed Forward Neural Network architecture with 2 hidden layers, the first with 128 nodes and the second with 64 nodes. The input layer accepts 6 parameters ($\Omega_{\rm DM}$, $\log_{10} A_s$, $f_{\rm ax}$, $\log_{10}m_{\rm ax}$, $z$ and $k$), and outputs a single value: the prediction for the power-spectrum ratio for the given input values. We find that using the Gaussian Error Linear Unit (GELU) as activation functions and using the \textsc{L1loss} function to calculate the loss is sufficient. We train with learning rate $\gamma = 0.01$, weight decay set to $\beta = 5 \cdot 10^{-6}$ and use the Adam optimizer. Training is stopped once there is no significant improvement to the loss for $30$ epochs.




The data-set used for training is sampled using the LHS shown in \cref{fig:LHS}, where each point is a combination of parameters that we use to calculate the power-spectrum ratio. In the end, this data-set contains approximately $1.9 \cdot 10^{6}$ data-points. Depending on the chosen batch size, the training may take between $\sim$3 CPU hours and $\sim$15 CPU minutes to train. 
After training, the look-up for one set of $256$ $P_{\rm ax}(k)/P(k)$ values takes about $\sim 0.05$ s, which is a speed-up of a factor $\sim 4500$ from running the ${\it N}$-body simulation setup to produce the same result, which takes $\sim 4$ minutes. 

\begin{figure*}
    \centering
    \includegraphics[width=0.9\linewidth]{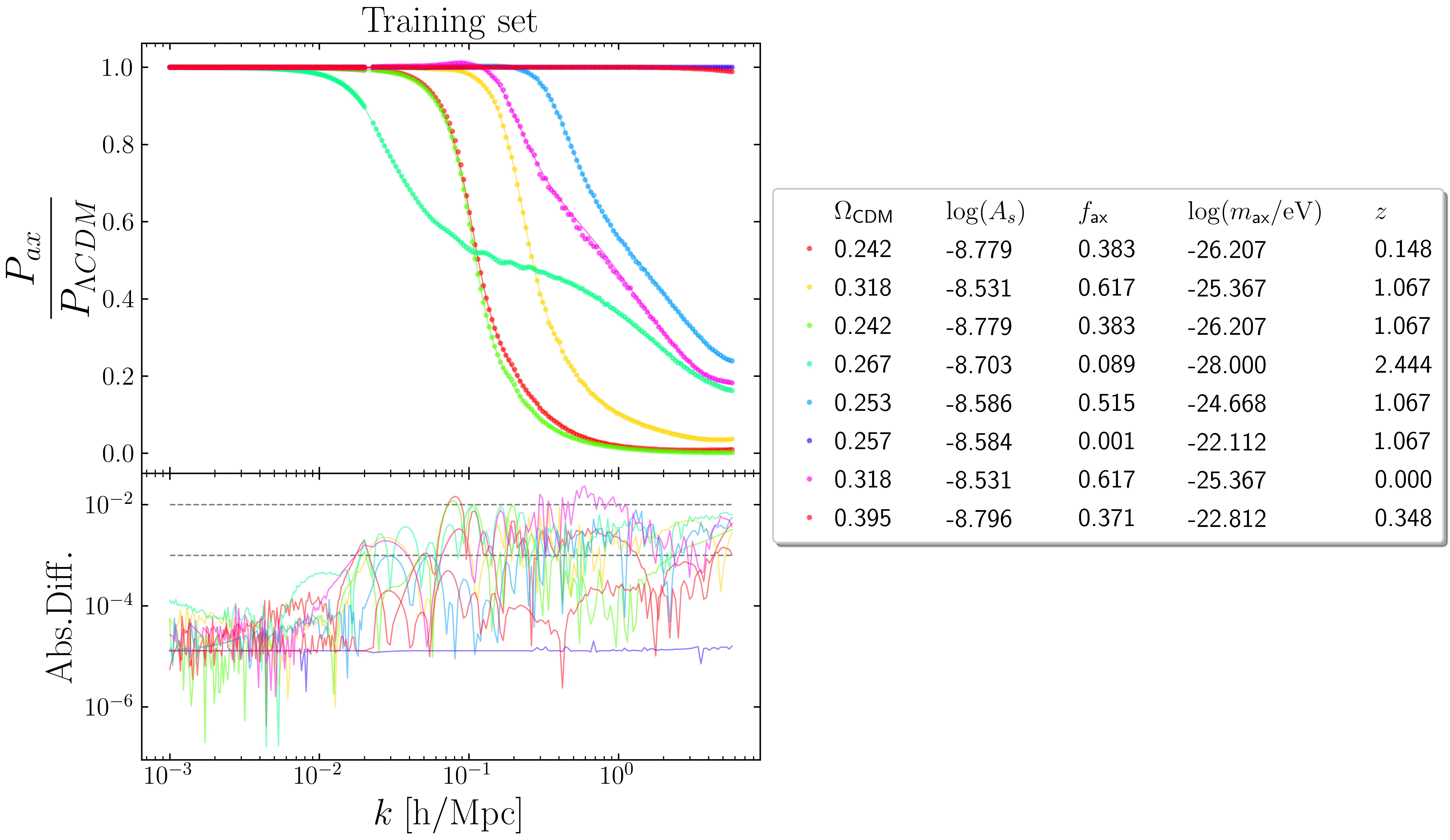}
    \includegraphics[width=0.9\linewidth]{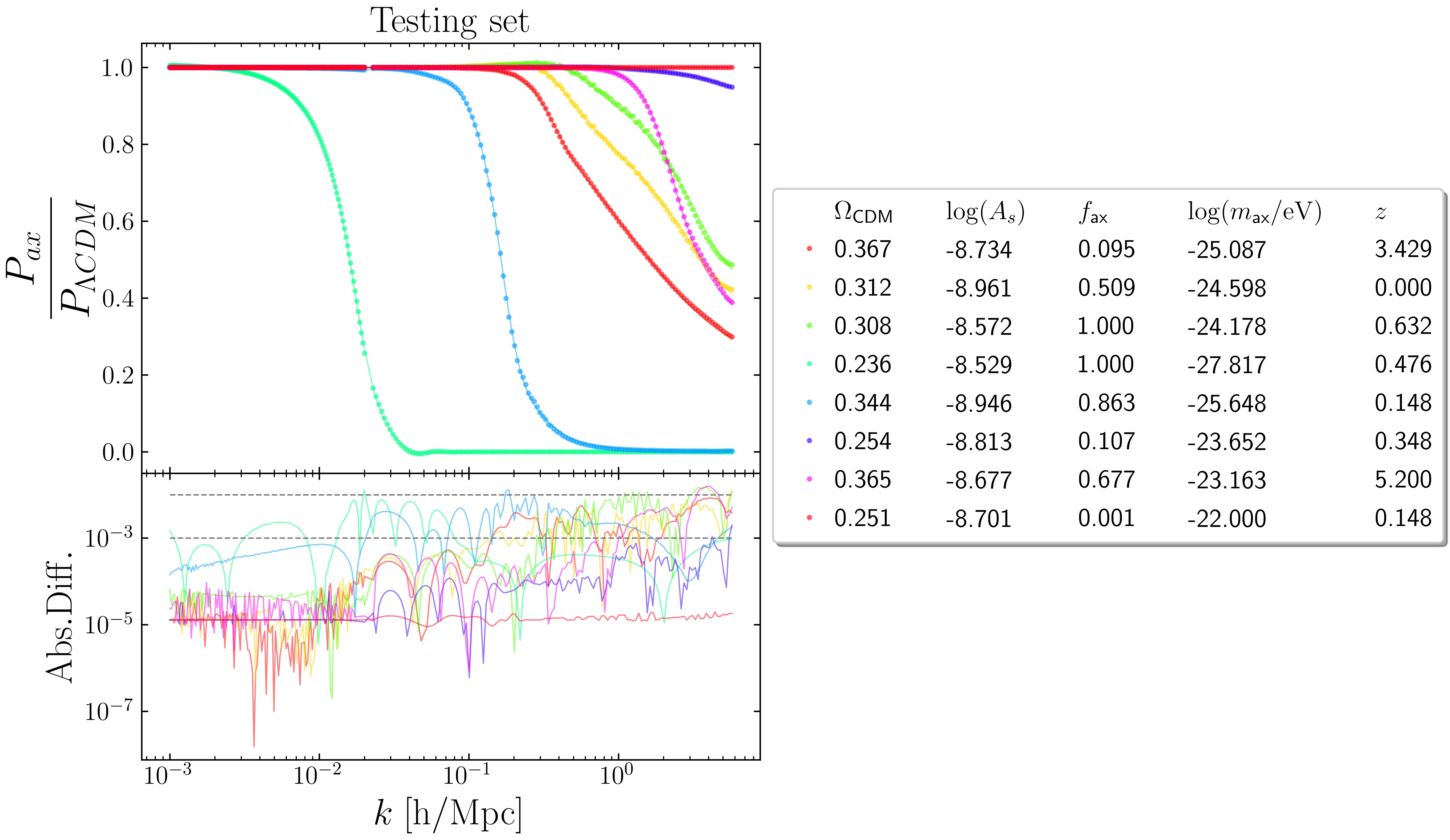}
    \caption{The emulator's performance on the training data-set (top) and test-set (bottom). The top panel shows the emulators prediction compared to the data from the simulations, and the bottom panel shows the squared error between the two. Solid lines indicate the prediction of the emulator, while dots indicate the data from simulations, with each color specifying a different set of cosmological parameters.}
    \label{fig:t_set}
\end{figure*}


\subsection{Emulator performance}

We test the accuracy of the emulator by comparing the emulator's prediction with the results from simulations, using the data-set used during training, and an independent data-set that the network has not been trained on. This comparison is shown in \cref{fig:t_set}, where the top panel compares the training data-set and the bottom panel compares the test data-set over $k$ for $8$ different combinations of the cosmological parameters, with error measured in squared difference:
\begin{equation}
    E(x_i, y_i) = |x_i^2 - y_i^2|.
\end{equation}
While the error varies between samples, we find that the overall error lies below $10^{-2}$. In some cases, we find a jump in error above $10^{-2}$, which may be explained by a slight misalignment between the \textsc{axionCAMB} data and the data from \textsc{COLA}. This may give rise to a slightly higher error in the emulator, as these transitions get smoothed over during training. Another possible source could be noise in the power-spectrum boost calculated by \textsc{COLA}. 

\begin{figure}
    \centering
    \includegraphics[width=\linewidth]{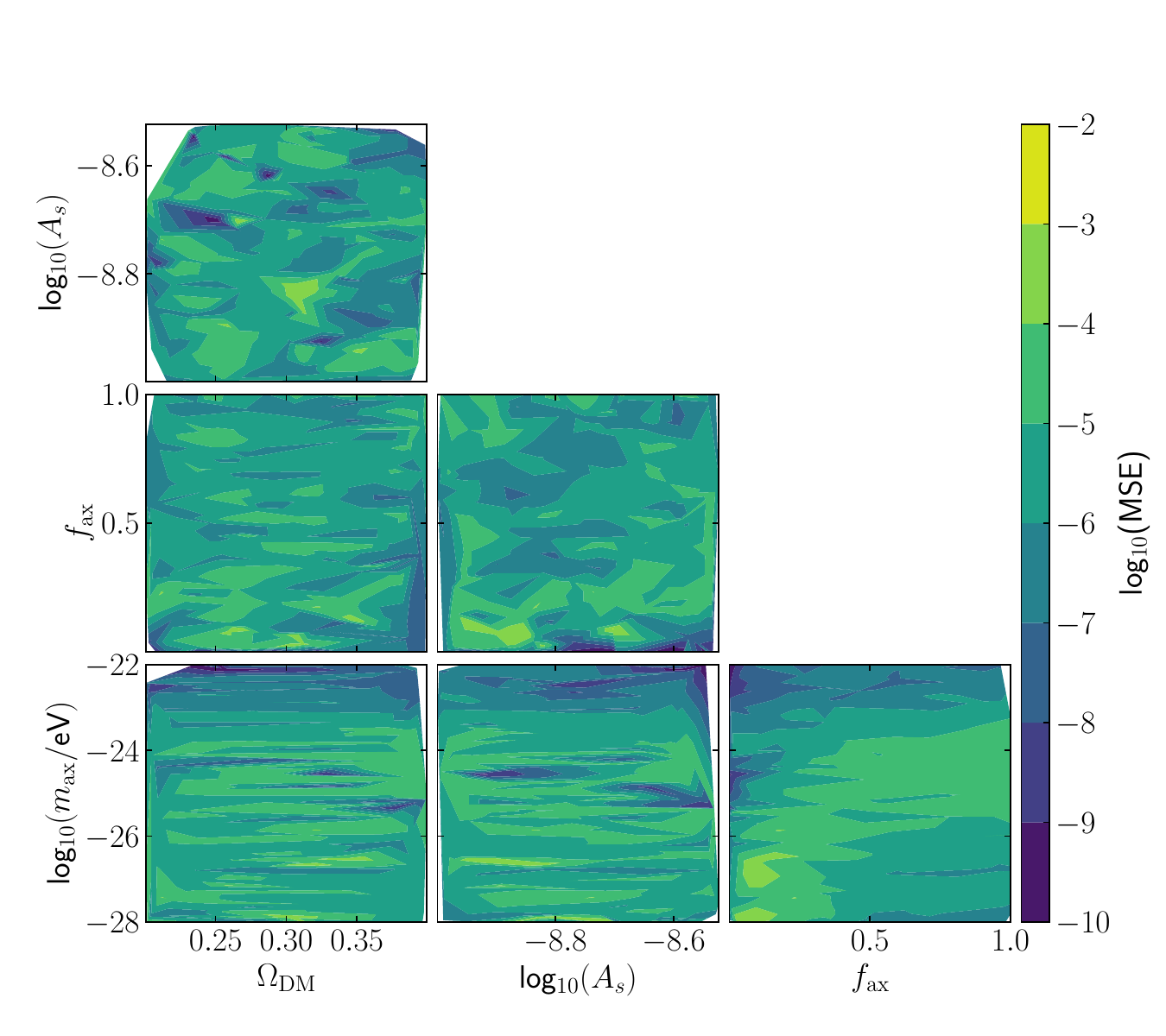}
    \caption{The mean squared error between the emulator and \textsc{COLA} at each point in the Latin hypercube sampling shown in figure \ref{fig:LHS}. We find a peak error of $10^{-2}$ for combinations of small axion abundances and small axion masses.}
    \label{fig:accuracy}
\end{figure}

To test the emulator's performance for different cosmological parameters, we calculate the mean squared error between the \textsc{COLA} simulations and the emulator's prediction for each data-point in \cref{fig:LHS}. The result is shown in \cref{fig:accuracy}. Notably, for combinations of small axion abundance and axion mass, there is an increase error, which may be due to fast changes in the power-spectrum boost, caused by the strong suppression of low mass axion cosmologies. 

\begin{figure}
    \centering
    \includegraphics[width=\linewidth]{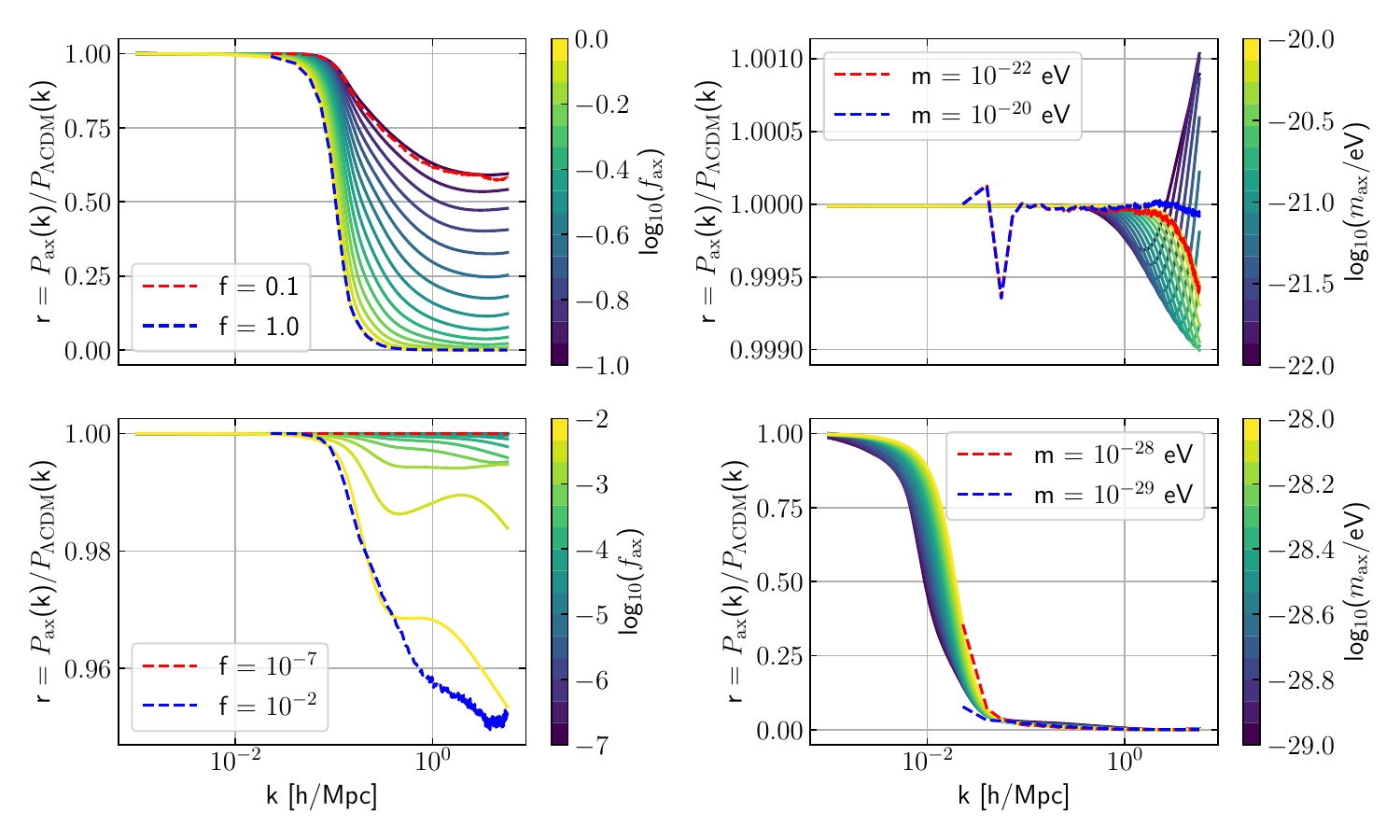}
    \caption{Emulator's performance close to and outside the parameter-space edges. The dashed lines indicate the power-spectrum ratio calculated using \textsc{COLA}. Comparing these results to that of the emulator, it is apparent that the emulator works well close to and just outside the parameter-space edges.}
    \label{fig:emu_lim}
\end{figure}

Additionally, we test the emulator's ability to generate extrapolated features beyond the trained parameter-space. In \cref{fig:emu_lim} we demonstrate how the emulator performs on untrained values of $m_{\rm ax}$ and $f_{\rm ax}$, and compare to the prediction from \textsc{COLA} simulations. We find that the emulator works well in the upper limits of $f_{\rm ax}$, that is $f_{\rm ax} \sim 1$, and in the chosen lower limits of $m_{\rm ax}$, when $m_{\rm ax} \sim 10^{-28}$ eV. In these cases, the emulator successfully predicts increased suppression for higher values of $f_{\rm ax}$ and suppression on larger scales for smaller $m_{\rm ax}$.
In the lower limits of $f_{\rm ax}$, we find that the emulator correctly predicts decreased suppression of power, while in the upper limit of $m_{\rm ax}$ the emulator encounters some discrepancies, as it predicts an increase in power of $\sim 1\%$. 

\section{Comparison}

\begin{figure*}
    \centering
\includegraphics[width=0.6\linewidth]{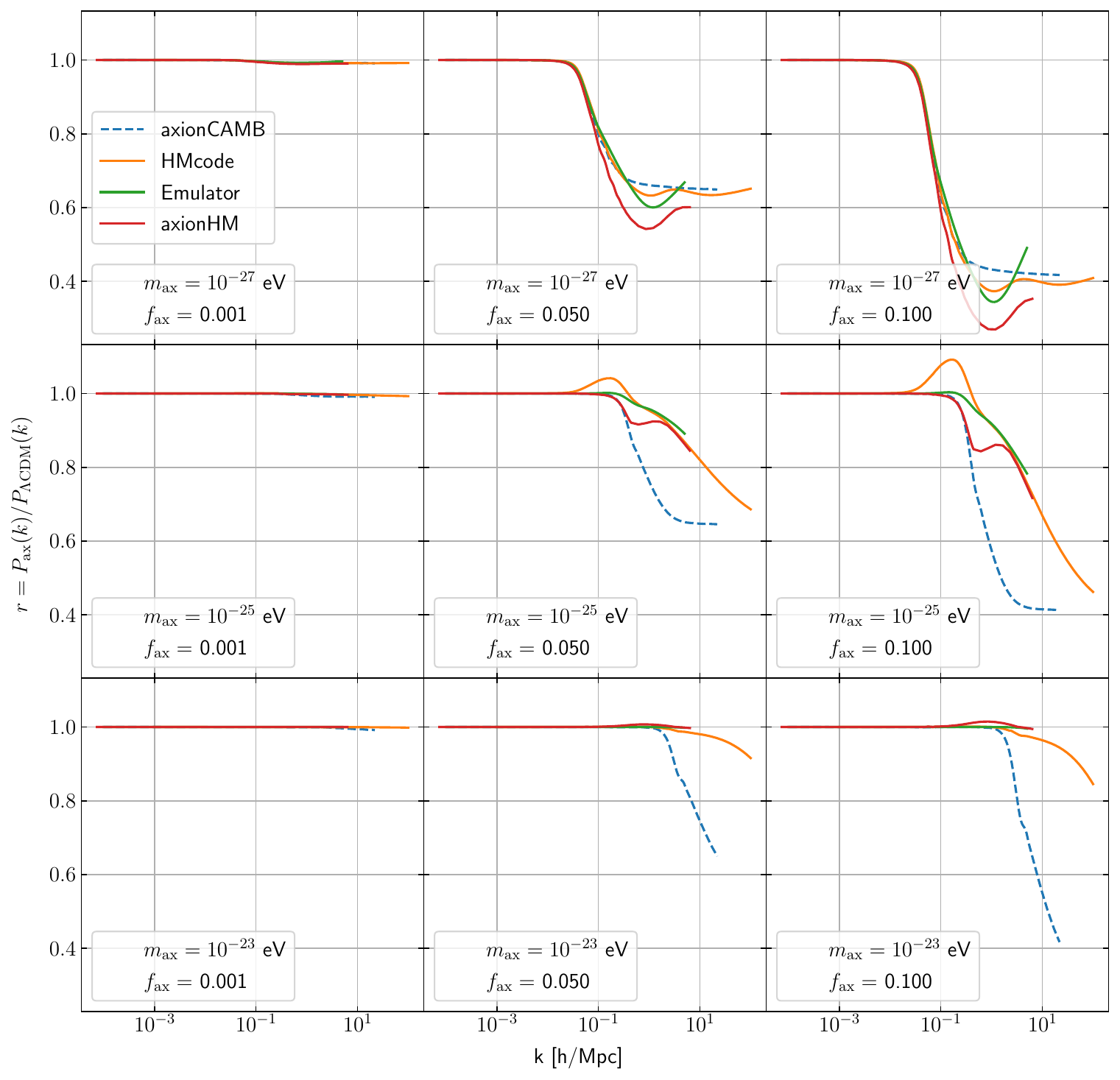}
\includegraphics[width=0.6\linewidth]{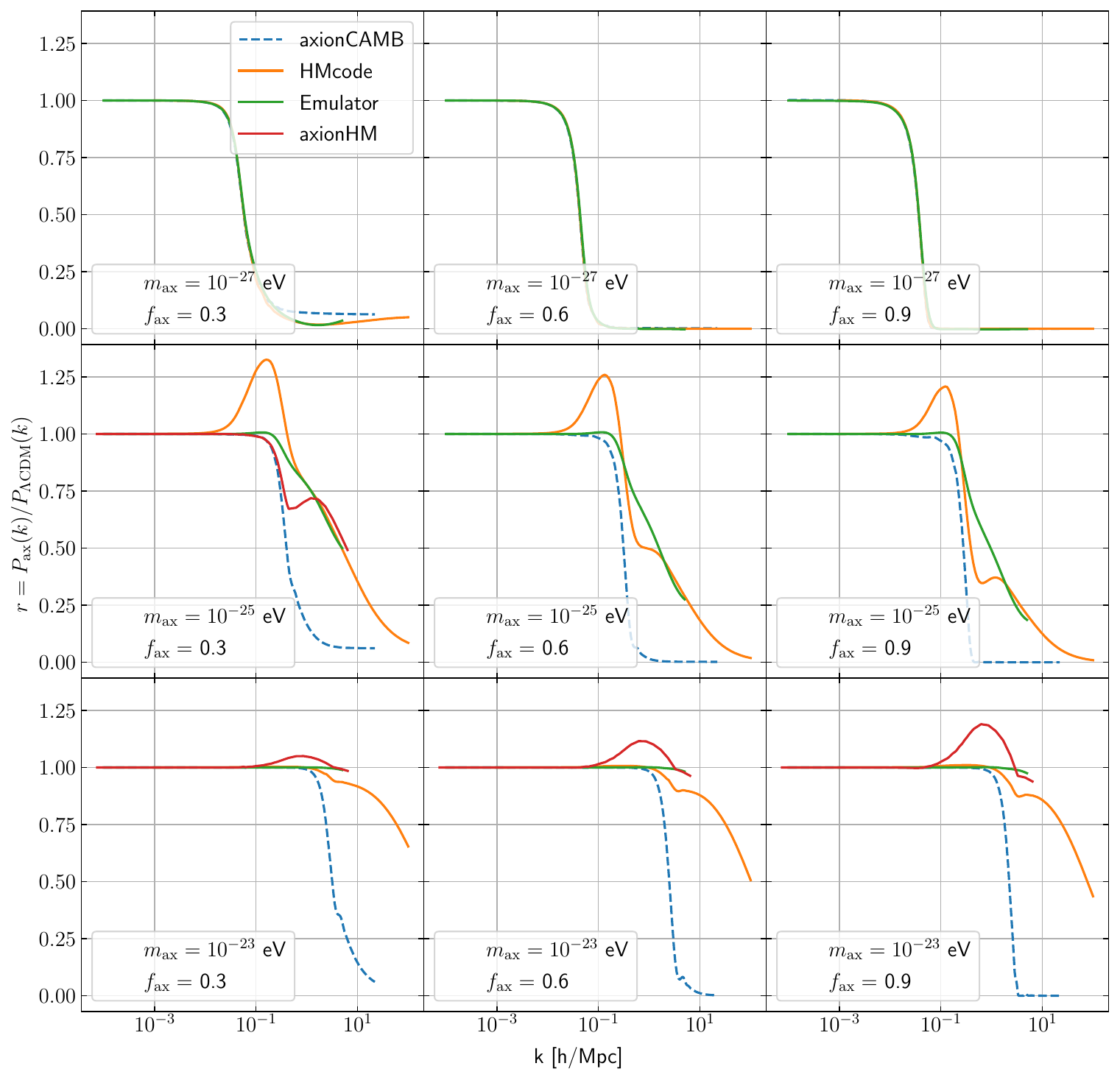}
    \caption{Comparison of the power-spectrum ratio from our emulator to that of naive \textsc{HMCode} within \textsc{CAMB}, \textsc{axionCAMB} (the linear prediction) and \textsc{axionHMCode} (a modification of \textsc{HMCode} including axion physics). We were unable to run \textsc{axionHMCode} for certain parameters choices (typically for large $f_{\rm ax}$).}
    \label{fig:comparison}
\end{figure*}

We compare the power-spectrum ratio predicted by the emulator with the linear prediction from \textsc{axionCAMB}, along with the standard non-linear prescription of \textsc{HMCode}\footnote{https://github.com/alexander-mead/HMcode} and \textsc{axionHMCode} \citep{Vogt_2023}. 
This comparison is shown in \cref{fig:comparison}, where the top $3\times3$ panels make the comparison for small axion abundances, and the bottom $3\times 3$ panels does so for larger abundances. Since \textsc{HMcode} considers CDM evolution and modified initial conditions, it should, in principle, provide the best match to our simulations, which also evolve pure CDM. 
However, because \textsc{HMCode} is calibrated to $\Lambda$CDM, it is not immediately clear how closely our simulations will follow its predictions. 
Meanwhile, \textsc{axionHMcode} improves on \textsc{HMcode} by including relevant axion physics, such as the impact of a cored density profile, in order to capture wave-like effects. Notably, \textsc{HMcode} does not explicitly model axion physics beyond using the axion linear power spectrum, so it is only expected a priori to be accurate for very small axion fractions. 
Due to restrictions in \textsc{axionHMcode}, our comparison is limited to certain combinations of $m_{\rm ax}$ and $f_{\rm ax}$. 



Overall, we find that the prediction from the emulator agrees with \textsc{axionHMcode} and \textsc{HMcode}. However, each approach predicts 
a slightly different degree of suppression and enhancement to the power-spectrum on scales smaller than the Jeans scale. 
For example, \textsc{HMcode} predicts an enhancement to the power-spectrum of up to $\sim 30\%$ for axion masses around $10^{-23}$ eV. On the other hand, \textsc{axionHMcode} predicts significant enhancements when $m_{\rm ax} \sim 10^{-23}$ eV and the axion abundance $f_{\rm ax} \gtrsim 0.1$. 
Both \textsc{HMcode} and \textsc{axionHMcode} also predict a spoon shape in the power-spectrum ratio on top of the suppression. Since \textsc{axionCAMB} only considers linear perturbations, it is not able to replicate this feature. This spoon shape can also be found when computing the ratio of power-spectrums of massive neutrinos to $\Lambda$CDM, using either simulations \citep{Brandbyge:2008} or halo-models \citep{Hannestad:2020}. The emulator does not predict any enhancements or significant spoon-shape.

\begin{figure*}
    \centering
    \includegraphics[width=\linewidth]{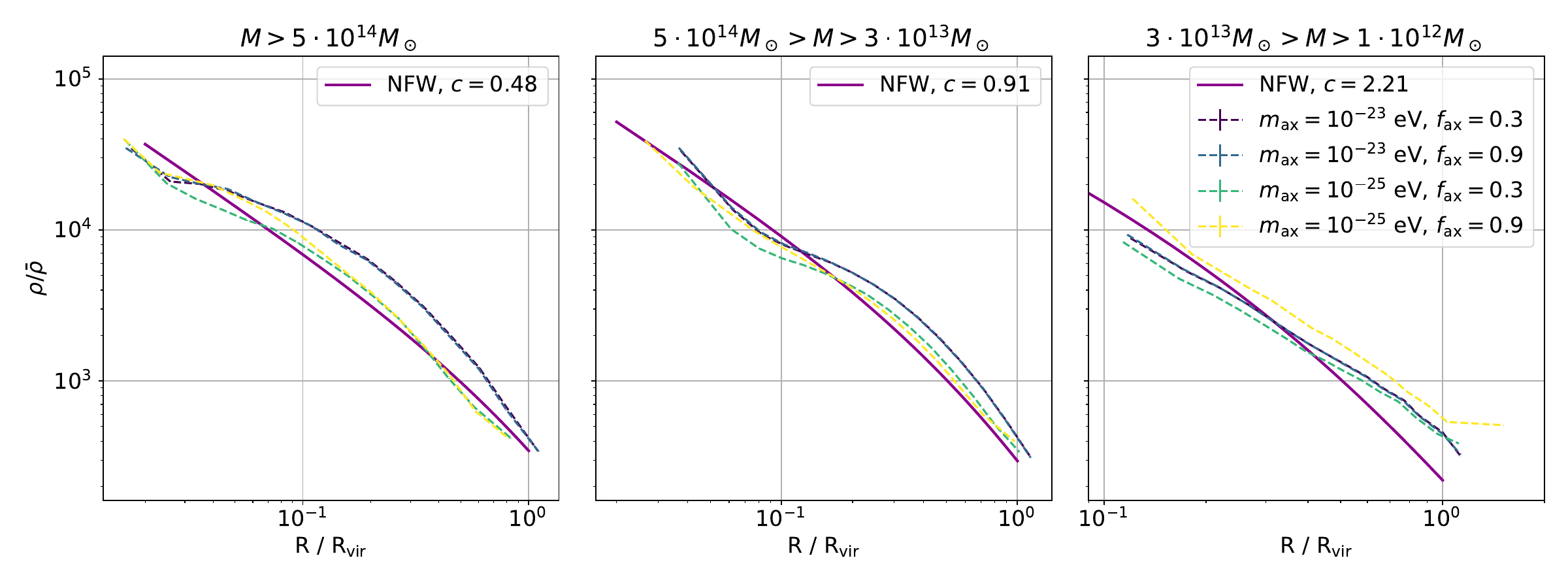}
    \caption{Calculated and fitted theoretical halo profiles at redshift $z=0$ for different combinations of $m_{\rm ax}$ and $f_{\rm ax}$, binned to three mass ranges.}
    \label{fig:halo_profile}
\end{figure*}
\begin{figure}
    \centering
    \includegraphics[width=\linewidth]{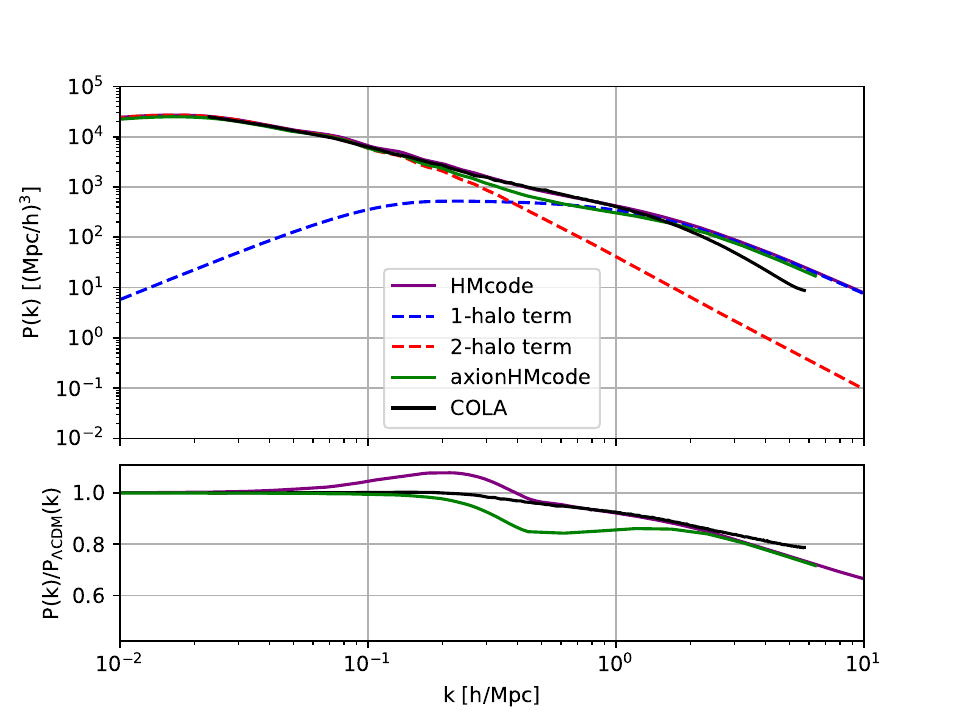}
    \caption{$m_{\rm ax} = 10^{-25}$ eV, $f_{\rm ax} = 0.1$. This shows the one and two halo terms together with the boost. We see that the differences between \textsc{HMCode}, \textsc{axionHMCode} and the emulator mainly comes from the transition region between the two regimes.}
    \label{fig:halofit}
\end{figure}

In the work done by \cite{Vogt_2023}, the enhancements to the power-spectrum are explained by the transition between the one- and two-halo terms overlapping with the soliton core in the halos, which dominates at that scale for axions with mass $\sim 10^{-22}$ eV. Similar enhancements have also been observed in simulations \cite{Nori:2019}. We find that enhancements can also be found when using \textsc{HMCode} to calculate the power-spectrum, but to a higher degree.
This discrepancy likely arises from the fact that \textsc{HMCode} is fitted to $\Lambda$CDM, and thus does not consider soliton cores in the 1-halo term.
\cref{fig:halofit}, which shows the 1- and 2-halo terms, highlights how the enhancements arises in the transition region.
Given the resolution used in this paper, we are not able to resolve the inner parts of the halos, where the halo profiles deviates from $\Lambda$CDM and instead takes the form of a soliton core \citep{Vogt_2023}, and thus are unable to reproduce any enhancements to the power-spectrum. At the same time, we do not expect to see a soliton core within the halos due to the lack of quantum interactions in the simulations. This means that, as seen in  \cref{fig:halo_profile}, our simulations mainly predicts NFW profiles even for very large axion fractions.

\begin{figure}
    \centering
    \includegraphics[width=\linewidth]{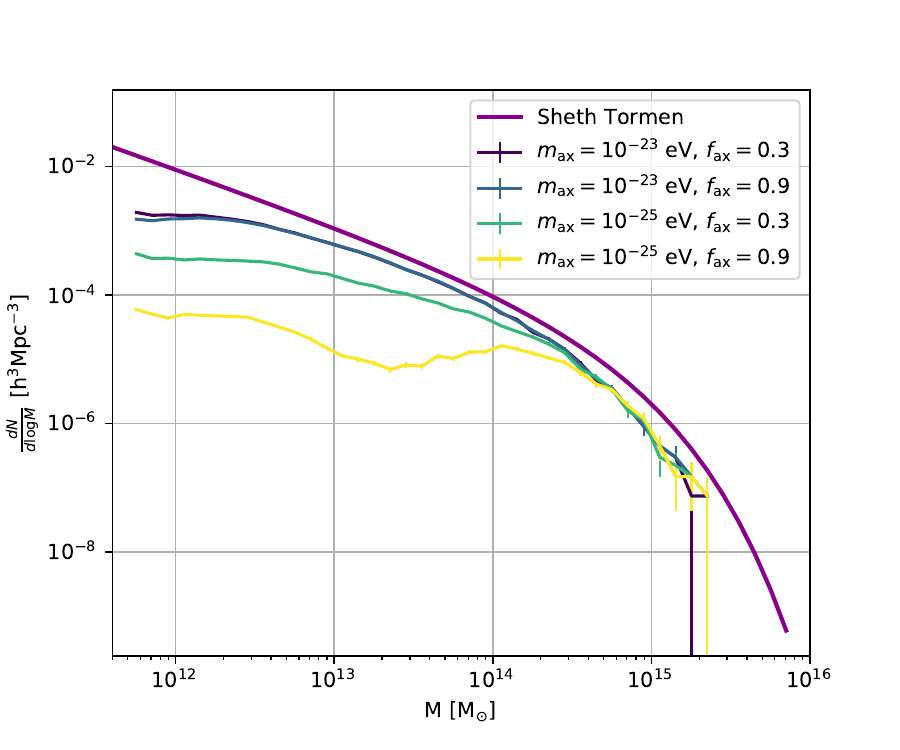}
    \caption{Theoretical and calculated halo mass function for different combinations of $m_{\rm ax}$ and $f_{\rm ax}$ at redshift $z=0$}
    \label{fig:halo_mass_function}
\end{figure}

The halo mass function in a fuzzy DM cosmology is affected by a cut-off for small mass halos \citep{Mihir:2022}. Most studies have found that at redshift $z=0$, axions with mass $10^{-22}$ eV will result in a cut-off in the halo mass function approximately between $10^8 M_\odot - 10^9 M_\odot$ (\cite{Marsh:2014}, \cite{Bozek:2015}, \cite{Du:2017}). Given the resolution used in this study, we do not have the power to resolve halos down to this mass. 
For smaller axion masses, \cite{Du:2017} found that the cut-off happened at increasingly higher masses, going as far as to $m_{\rm ax} = 10^{-24}$ eV, where they found a cut-off roughly around $\sim 10^{12}M_\odot$. For even lower masses, one would expect the cut-off to happen at even higher masses, which we are able to resolve given our simulation setup. As seen in \cref{fig:halo_mass_function}, while we are able to predict some suppression to the halo mass function with our approach, we are not able to reproduce the cut-off seen in the mentioned studies. Within the resolved mass range, we find that the halo mass function is reduced to the Sheth-Tormen approximation for large axion masses ($\sim 10^{-22}$ eV), and for the most massive halos. 


\section{Conclusion}

In this paper, we follow the approach of simulating a mixed axion and cold dark matter cosmology using an approximate $N$-body simulation, known as the \textsc{COLA} method, where the axion physics come into play only through the initial conditions. While this approach is much faster than simulating the axions by solving the Schr\"{o}dinger-Poisson equations, it suffers from lack of resolution on small scales, due to the approximative nature of the simulations, and lacks the quantum pressure associated with the axions. Some of the inaccuracies that come with a lack of resolution can be factored out by looking at the ratio between the power spectrum with axions to that of $\Lambda$CDM. This also factors out some of the dependency on the cosmological parameters. 

Using this approach, we create a data-set which is used to train an emulator which predicts the boost to the power-spectrum. The emulator takes six input parameters: the mass of the axion $m_{\rm ax}$, the axion abundance $f_{\rm ax}$, $A_s$, $\Omega_{\rm DM}$, redshift $z$, and the wave number $k$. 
We find that this emulator is in overall agreement with \textsc{axionHMcode}, which is a halo-based approach to calculate the matter power spectrum and includes the relevant axion physics. Compared to this approach, the emulator successfully predicts the suppression to the power spectrum, but fail to reproduce enhancements or a spoon-like shape seen in \textsc{axionHMcode}. Comparing to \textsc{HMcode} on the other hand, a code that is fitted to $\Lambda$CDM and does not have any axion physics included, the fit is much worse. Some of the discrepancy here is likely produced by the way the transition between the 1- and 2-halo terms is modeled. To determine the true accuracy of our method, a proper comparison to full axion simulations is required. However, such simulations are very expensive (especially for large box sizes) and were not publicly available at the present time. We hope to do this in the future. 



Since our approach simulates the axion particles purely as standard cold dark matter particles, certain features may be hard to reproduce. The quantum pressure of the axions alter the halo profiles, leading to a soliton core. This is a feature that we do not expect to replicate since we do not have the quantum pressure in our simulations, and rather find that the profiles follow an NFW profile. 
In a mixed dark matter cosmology, the number of low-mass halos are greatly suppressed, resulting in a cut-off in the halo mass function. We do not have the resolution to fully resolve these halos. Nevertheless, we do see some effect of the axions as lighter axion masses shows sign of increasing suppression in the halo mass function.

The total cost of our emulator, including simulations and training, was around $2000$ CPU hours. Given its speed and minimal computational requirements, our emulator shows promise as a powerful tool for inferring the axion mass and abundance from matter power spectrum observations. By combining our emulator with an emulator for $\Lambda$CDM one can quickly predict the power spectrum for mixed dark matter, and significantly reduce the computational load associated with running inference approaches such as MCMC samplings. 

The code and trained emulator is available on GitHub \url{https://github.com/frdennis/Axion-Emulator}.

\section{Acknowledgements}
We thank David Alonso for useful discussion and Cheng-Zong Ruan for sharing codes used for the emulation. We also thank David Marsh and Kier Rogers for sharing insightful ideas and providing valuable feedback.

\bibliographystyle{aa}
\bibliography{lib}

\end{document}